\documentclass[apsrev,superscriptaddress,amsmath,amsfonts,amssymb,floatfix,showpacs,showkeys]{revtex4}

\begin{document}
\title{Comment on ``Comment on ‘Supersymmetry, PT-symmetry and spectral bifurcation ’’}

\author{Kumar Abhinav}
\affiliation{Indian Institute of Science Education and Research-Kolkata, Mohanpur Campus, Nadia-741252, West Bengal, India}
\email{kumarabhinav@iiserkol.ac.in}

\author{P. K. Panigrahi}
\affiliation{Indian Institute of Science Education and Research-Kolkata, Mohanpur Campus, Nadia-741252, West Bengal, India}
\email{pprasanta@iiserkol.ac.in}

\begin{abstract} 
In ``Comment on Supersymmetry, PT-symmetry and
spectral bifurcation" \cite{BQ1},  Bagchi and  Quesne correctly show the presence of a class of states for the complex Scarf-II potential in the unbroken PT-symmetry regime, which were absent in \cite{AP}. However, in the spontaneously broken PT-symmetry case, their argument is incorrect since it fails to implement the condition for the potential to be PT-symmetric: $C^{PT}[2(A-B)+\alpha]=0$. It needs to be emphasized that in the models considered in \cite{AP}, PT is spontaneously broken, implying that the potential is 
PT- symmetric, whereas the ground state is not. Furthermore, our supersymmetry (SUSY)-based 'spectral bifurcation' holds \textit{independent} of the $sl(2)$ symmetry consideration for a large class of PT-symmetric potentials. 

\end{abstract}



\maketitle

The primary goal of the paper ``Supersymmetry, PT-symmetry and
spectral bifurcation'' \cite{AP} was to analyze the condition for spontaneous PT-symmetry breaking for a wide class of potentials. 
\\
The condition for the complex Scarf-II potential \cite{Ahmed} to be PT-symmetric, under suitable parameterization, came out to be, 

\begin{equation}
C^{PT}[2(A-B)+\alpha]=0.  
\end{equation}

In the unbroken PT-symmetry regime, it was found that $C^{PT}=0$, and the corresponding superpotential was,

\begin{equation}
W(x)=Atanh(\alpha x)+iBsech(\alpha x), 
\end{equation}

yielding the potential,

\begin{equation}
V_{-}(x)=-\left[A(A+\alpha)+B^2\right]sech^{2}(\alpha x)+iB(2A+\alpha)sech(\alpha x)tanh(\alpha x).
\end{equation}

When PT-symmetry is spontaneously broken, $C^{PT}\neq 0$, meaning $A=B-\frac{\alpha}{2}$. This results in a unique potential,

\begin{equation}
 V_{-}(x)=-\left[2A(A+\alpha)-2(C^{PT})^{2}+\frac{{\alpha}^2}{4}\right]sech^{2}(\alpha x)+i\left[2A(A+\alpha)+2(C^{PT})^{2}+\frac{{\alpha}^{2}}{2}\right]sech(\alpha x)tanh(\alpha x),
\end{equation}

corresponding to two \textit{different} superpotentials,

\begin{equation}
 W^{\pm}(x)=\left(A\pm iC^{PT}\right)tanh(\alpha x)+\left[\pm C^{PT}+i\left(A+\frac{\alpha}{2}\right)\right]sech(\alpha x),
\end{equation}

representing two disjoint sectors of the Hilbert space with normalizable wave-functions.
\\
 We agree with Bagchi and Quesne that in the unbroken PT-symmetry regime, a further symmetry in the parameter space yields another normalizable set of wavefunctions having different spectrum, which owes its origin to an underlying $sl(2)$ symmetry \cite{BQ2}.
\\

Bagchi and Quesne \cite{BQ1} further demonstrate that, when PT-Symmetry is spontaneously broken, the $sl(2)$ symmetry of the potential is realized through the exchange, 

\begin{equation}
\mathcal{A}+\frac{\alpha}{2}\leftrightarrow \mathcal{B},
\end{equation}

 where $\mathcal{A}=A\pm iC^{PT}$ and $\mathcal{B}=B\mp iC^{PT}$, resulting yet again in two disjoint sectors in the Hilbert space. The corresponding ground-state energies are $-{\mathcal{A}}^2$ and $-{\mathcal{B}}^2$. It needs to be emphasized that in the models considered in \cite{AP}, PT is spontaneously broken, implying that the potential is PT- symmetric, whereas the ground state is not. Hence, the condition given in Eq.(1), holds in both the broken and unbroken sectors. For spontaneously broken PT-symmetry, we have $C^{PT}\neq 0$ \cite{AP}. This yields $A+\frac{\alpha}{2}=B$, which reduces the parametric $sl(2)$ exchange to,
 
 \begin{equation}
C^{PT}\leftrightarrow -C^{PT}.
\end{equation}

Then, the ground state energies of both the sectors turn out to be $-(A\pm iC^{PT})^2$. Therefore, the $sl(2)$ transformation, when PT-symmetry is broken, merely relates the bifurcated sectors of the Hilbert space, obtained already through the application of SUSY \cite{AP}. Hence, both the sectors of the Hilbert space for unbroken PT-symmetry, under $sl(2)$ symmetry, maps to the same pair of sectors when PT-symmetry is spontaneously broken. This is the reason why, despite overlooking the $sl(2)$ algebra, the present authors obtained the complete complex-conjugate spectra. Furthermore, it was correctly found out in \cite{AP} that $C^{PT}\neq 0$ is the sole parametric criterion for broken PT-symmetry, resulting in the spectral bifurcation.\\ \\
Our method was applied to a number of other potentials, tabulated in \cite{AP}, which do not satisfy the $sl(2)$ algebra. In each case, the spectral bifurcation was present for $C^{PT}\neq 0$. This further shows that Bagchi and Quesne's approach does not lead to the SUSY-parametric criterion of spontaneous breaking of PT-symmetry. Further, the two superpotentials when PT-symmetry is preserved, maps \textit{independently} to the same pair of superpotentials when PT-symmetry is broken, which is not clear in \cite{BQ1, BQ2}.\\ \\


\textbf{References}
\begin{thebibliography}{99}

\bibitem{BQ1} B. Bagchi, C. Quesne, arXive, quant-ph (2010).
\bibitem{AP} K. Abhinav, P.K. Panigrahi, Ann. Phys. 325 (2010) 1198.
\bibitem{BQ2} B. Bagchi, C. Quesne, Phys. Lett. A 273 (2000) 285.
\bibitem{Ahmed} Z. Ahmed, Phys. Lett. A, 282 (2001) 343-348.
\end{thebibliography}
\end{document}